\documentstyle[11pt,aas2pp4]{article}
\input psfig
\psfull
\def\be{\begin{equation}}
\def\ee{\end{equation}}
\def\ba{\begin{eqnarray}}
\def\ea{\end{eqnarray}}
\def\wisk#1 {\ifmmode{#1}\else{$#1$}\fi}
\def\ob    {\Omega_{b}}

\def\Q      {Q_{10}}

\def\ol    {\lambda_{o}}
\def\oo    {\Omega_{o}}

\def\oh    {\Omega_{HDM}}
\def\obh  {\Omega_{b}h^{2}}
\def\etal   {{\sl et al.}~\rm}
\def\apj    {{\sl Ap.J.$\!$}~\rm}

\newcommand {\lsim}{\mbox{$\:\stackrel{<}{_{\sim}}\:$} }

\newcommand {\gsim}{\mbox{$\:\stackrel{>}{_{\sim}}\:$} }
\def\ap     {A_{peak}}
\def\lp     {\ell_{peak}}
\def\free   {$*^{e}$}
\begin{document}
\title{What Can Cosmic Microwave Background Observations\\
 Already Say About Cosmological Parameters\\
in Open and Critical-Density Cold Dark Matter Models?
\footnote{\apj, 496, in press (April 1, 1998)}}
\author{ Charles H. Lineweaver 
\footnote{School of Physics, UNSW, Sydney 2052, Australia,  charley@bat.phys.unsw.edu.au}
$^{\! ,}$\footnote{Observatoire de Strasbourg, 11 rue de l'Universit\'e, 67000 Strasbourg, France}
and Domingos Barbosa $^{3,\! }$
\footnote{Centro de Astrof\'{\i}sica da U.P., Rua do Campo Alegre 823, 4150 Porto, Portugal}}

%%******************************************************************
\begin{abstract}
We use a combination of the most recent cosmic microwave background (CMB) 
flat-band power measurements to place constraints on Hubble's constant $h$  
and the total density of the Universe $\oo$ in the context of inflation-based
cold dark matter (CDM) models with no cosmological constant.
We use $\chi^{2}$ minimization to explore the 4-dimensional parameter 
space having  as free parameters, $h$, $\oo$, the power spectrum slope $n$ and
the power spectrum normalization at $\ell = 10$.
Conditioning on $\oo = 1$ we obtain $h= 0.33\pm 0.08$.
Allowing $\oo$ to be a free parameter reduces the ability of the
CMB data to constrain $h$ and we obtain $0.26 < h < 0.97$  
with a best-fit value at $h=0.40$. 
%$h= 0.35^{+0.63}_{-0.09}$.
We obtain $\oo = 0.85$ and set a lower limit $\oo > 0.53$.
A strong correlation between acceptable
$h$ and $\oo$ values leads to a new constraint
$\oo h^{1/2}= 0.55 \pm 0.10$.
We quote $\Delta \chi^{2} = 1$ contours as error bars, however because
of nonlinearities of the models, these may be only crude approximations
to $1 \sigma$ confidence limits.

A favored open model with $\oo = 0.3$ and
$h = 0.70$ is more than $\sim 4 \sigma$ from the CMB data
best-fit model and is rejected by goodness-of-fit statistics
at the $99\%$ CL.
High baryonic models ($\obh \sim 0.026$) yield the best CMB $\chi^{2}$ fits
and are more consistent with other cosmological constraints.
The best-fit model has
$n=0.91^{+0.29}_{-0.09}$ and $\Q= 18.0^{+1.2}_{-1.5}\;\mu$K.
Conditioning on $n=1$ we obtain $h = 0.55^{+0.13}_{-0.19}$,
$\oo = 0.70$ with a lower limit $\oo > 0.58$ and $\Q= 18.0^{+1.4}_{-1.5}\:\mu$K.
The amplitude and position of the dominant peak in the best-fit
power spectrum are $\ap = 76^{+3}_{-7}\;\mu$K and $\lp = 260^{+30}_{-20}$.

Unlike the $\oo = 1$ case we considered previously, 
CMB $h$ results are now consistent with the higher
values favored by local measurements of $h$ but only if
$0.55 \lsim \oo \lsim 0.85$.
Using an approximate joint likelihood to combine our 
CMB constraint on $\oo h^{1/2}$ with other cosmological constraints 
we obtain $h=0.58 \pm 0.11$ and $\oo = 0.65^{+0.16}_{-0.15}$.

\end{abstract}

%%%%%%%%%%%%%%%%%%%%%%%%%%%%%%%%%%%%%%%%%%%%%%%%%%%%%%%%%%%%%%%%%%%
\section{INTRODUCTION}
\label{sec:intro}
%%%%%%%%%%%%%%%%%%%%%%%%%%%%%%%%%%%%%%%%%%%%%%%%%%%%%%%%%%%%%%%%%%%

The ensemble of cosmological data prefers
best-bet universes which seem to congregate in several distinct regions of 
parameter space (Ostriker \& Steinhardt 1995, Viana 1996).
Among the best-bet universes, open models figure prominently and
are possibly the favorite candidate (Liddle \etal 1996a). 
This preference is mainly due to observational evidence
(e.g., Willick \etal 1997, Carlberg \etal 1996, Dekel 1997).
Further motivation for examining $\oo < 1$ models is that
galaxy cluster baryonic fraction limits seem to be inconsistent with 
Big Bang nucleosynthesis (BBN) if $\oo = 1$ and 
$h \ga 0.50$ ($h=H_{o}/ 100\; km\: s^{-1} Mpc^{-1}$).
This ``baryon catastrophe'' has led some to believe 
that $\oo < 1$.

Recently, theoretical open universe models have been developed. 
Open-bubble inflation models have been developed by
Ratra \& Peebles (1994), Bucher, Goldhaber \& Turok (1995),
Yamamoto, Sasaki \& Tanaka (1995). Open hybrid inflation has
also been considered  (Garc\'{\i}a-Bellido \& Linde 1997).

\subsection{What Kind of Open Models We Consider and Why}

CMB measurements have become sensitive enough to constrain cosmological 
parameters in restricted classes of models.
In Lineweaver \etal (1997), (henceforth ``paper 1''), we described
our $\chi^{2}$ method and compared CMB data to predictions of 
COBE-normalized critical-density 
universes with Harrison-Zel'dovich ($n=1$) power spectra.
We briefly looked at CDM and flat $\Lambda$ CDM models by
exploring the $h - \ob$ plane and  the $h - \ol$ plane.
We used predominantly goodness-of-fit statistics to locate the regions of 
parameter space preferred by the CMB data.

In Lineweaver \& Barbosa (1998), (henceforth ``paper 2''), we used a similar
technique, again in critical-density universes, to explore the 4-dimensional 
parameter space $h$, $\ob$, $n$ and $Q$.
We obtained the result that if $\oo = 1$ (and our other assumptions are correct)
then the CMB data prefer surprisingly low values of the Hubble constant: $h \approx 0.30$.
We found that four independent
cosmological constraints also favored these low values in the $\oo = 1$ models considered.
This is in contrast to local measurements of $h$ which 
seem to prefer $h\approx 0.65 \pm 0.15$ (Freedman 1998, Tammann \& Federspiel 1997).

The $\oo = 1$ assumption we have made in our previous analyses can be considered very 
restrictive since plausible values for
$\oo$ in the range $0.2 \lsim \oo \lsim  1.0$ can change the power spectrum significantly.
In this work we consider open models motivated by the question:
Does our $h \approx 0.30$ result depend on the fact that we
limited ourselves to $\oo = 1$? 
Would a favored open model ($h = 0.7$ and $\oo = 0.3$)
be acceptable to the combined CMB data?
What pairs of $(h,\oo)$ values are compatible with the CMB data?

There are reasons to believe that $\oo < 1$ models will allow higher
$h$ values.
In paper 2 we found that the position of the primary acoustic peak 
in the angular power spectrum
is a dominant feature determining the low value of $h$. 
The position of the peak is shifted towards higher $\ell$ values in $\oo < 1$ models 
and this should have the effect of increasing the $h$ values of the best-fit models.
Motivated by this idea and the more general idea of increasing the size of the
parameter space into interesting regions, in this paper
we put constraints on the cosmological parameters
$h$, $\oo$, $n$, and the normalization at $\ell = 10$ in the context of $\oo \le 1$ CDM models.
We assume adiabatic initial conditions with no cosmological
constant.
As in paper 1 and 2, we take advantage of the recently available fast Boltzmann 
code to make the parameter-dependent model power spectra (Seljak and Zaldarriaga 1996).
We do not consider $\oo > 1$ models because the code 
is not yet available.

The recent dynamic interplay
between theory (providing a fast code to make model specific
predictions) and observations (new measurements are coming in about
 once a month)
is increasing our ability to distinguish models.
Major efforts have been and are being put into obtaining flat-band power
estimates.
The synthesis of these efforts is an important step towards a more complete picture
of the Universe.
Since the main goal of two new CMB satellites (MAP and Planck Surveyor)
is to constrain cosmological parameters,
it is important and timely to keep track of the
data's increasing ability to reject larger regions of parameter space
and put tighter constraints on preferred models.
That is the purpose of this paper.

Previous analyses most closely related to this work include Ganga \etal (1996),
White \& Silk (1996), White \etal (1996), Hancock \etal (1998), Bond \& Jaffe (1997), 
deBernardis \etal (1997).
Although methods, models and data sets differ, in the limited cases where
comparison is possible we have found no large discrepancies.

In Section 2 we summarize the method used to obtain the
results and examine some of the special features of open models.
In Section 3 we present our $h - \oo$ results and in Section
4 we compare them to non-CMB results.
In Section 5 we present our results for $n$, the normalization, $A_{peak}$ and
$\ell_{peak}$.
In Section 6 we discuss and summarize.

%%%%%%%%%%%%%%%%%%%%%%%%%%%%%%%%%%%%%%%%%%%%%%%%%%%%%%%%%%%%%%%%%%%
\section{METHOD}

%%%%%%%%%%%%%%%%%%%%%%%%%%%%%%%
\subsection{Data and $\chi^{2}$ Analysis}

We use a combination of the most recent CMB 
flat-band power measurements to place constraints on 
$h$, $\oo$, $n$ and the normalization at $\ell = 10$, $\Q$
(see Section 2.2 for $\Q$ definition).
We examine how the constraints on any one of these parameters changes 
as we  condition on as well as minimize with respect to the other parameters.
We  obtain best-fit values and approximate likelihood 
intervals for these parameters.

%%%%%% Figure 1 %%%%%%%%%%%%%%%%%
\begin{figure*}[t!]
\centerline{\psfig{figure=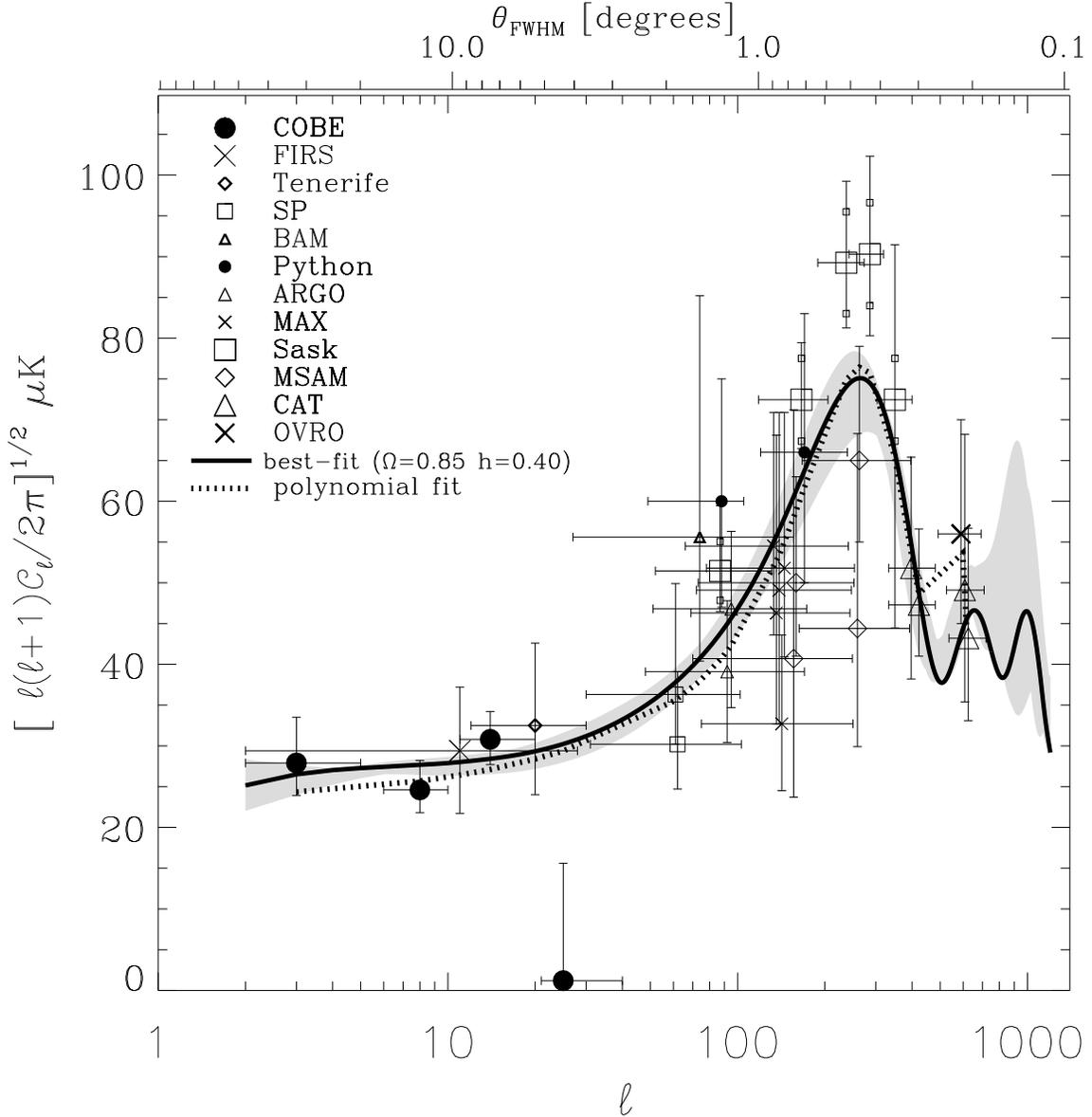,height=16.0cm,width=16.cm,bbllx=10pt,bblly=110pt,bburx=594pt,bbury=690pt}}
%\centerline{\psfig{figure=cl1alpeakh40.ps,height=18.0cm,width=18.cm,bbllx=10pt,bblly=110pt,bburx=594pt,bbury=690pt}}
%\centerline{\psfig{figure=cl1alpeak1sigmagrey.ps,height=18.0cm,width=18.cm,bbllx=10pt,bblly=110pt,bburx=594pt,bbury=690pt}}
%\centerline{\psfig{figure=cl1alpeak1sigma.ps,height=18.0cm,width=18.cm,bbllx=10pt,bblly=110pt,bburx=594pt,bbury=690pt}}
%\centerline{\psfig{figure=data.ps,height=18.0cm,width=18.cm,bbllx=10pt,bblly=100pt,bburx=594pt,bbury=690pt}}
%\centerline{\psfig{figure=hob1.ps,height=9cm,width=\hsize,bbllx=10pt,bblly=120pt,bburx=594pt,bbury=690pt}}
%%%BoundingBox: 28 70 594 636
%
\caption{Recent CMB observations compared with the best-fit model from 
Figure 4   %\protect\ref{fig:honfreeqfree} (solid line).
The dotted line is a sixth order polynomial fit to the data
which has a peak amplitude and position: $A_{peak} \approx 77\:\mu$K 
and $\ell_{peak} \approx 260$.
The grey region represents the $\sim 1\sigma$ contour in 
Figure 4;   %\protect\ref{fig:honfreeqfree}; 
that is, the power spectra from models within $\sim 1\sigma$
of the best-fit model are contained within the grey region.
The small squares above and below the 5 Saskatoon points represent the $7\%$ correlated
calibration uncertainty (Leitch 1998).
The best-fit model has $n=0.91$, $\Q=18.0\;\mu$K and $\obh = 0.026$.\label{data}}
\end{figure*}

We update the data of paper 2 to include several more points: \\
$\bullet$ updated Tenerife point (Guti\'errez \etal 1997):  
$\delta T_{eff} = 32.5^{+10.1}_{-8.5}\;\mu$K at $\ell_{eff} = 20$\\
$\bullet$ new MSAM results (Cheng \etal 1997):
$\delta T_{eff} = 50^{+13}_{-9}\;\mu$K at $\ell_{eff} = 159$ and
$\delta T_{eff} = 65^{+14}_{-10}\;\mu$K at $\ell_{eff} = 263$\\
$\bullet$ new preliminary CAT results  (Baker 1997):
$\delta T_{eff} = 47.3^{+9.3}_{-6.3}\;\mu$K at $\ell_{eff} = 422$ and
$\delta T_{eff} = 43.2^{+13.5}_{-10.1}\;\mu$K at $\ell_{eff} = 615$\\
$\bullet$ new preliminary OVRO result  (Leitch 1998):
$\delta T_{eff} = 56^{+14}_{-11}\;\mu$K at $\ell_{eff} = 537$.\\  %598$.\\

The current CMB flat-band power estimates used in this analysis are 
listed in Table 1 and
plotted in Figure 1.   %\ref{fig:data}.
Since there is much scatter in the data, there is much scepticism 
about the ability of the points to prefer any particular
region of parameter space.
We showed in papers 1 and 2 however that a simple $\chi^{2}$ analysis of interesting 
restricted families of models is capable of showing substantial preferences for
relatively small regions of parameter space.
The scatter in the data is partially deceiving in the sense that averaging
the data over broader bands in $\ell$ reduces the scatter and presents
a surprisingly coherent power spectrum which roughly follows the polynomial
fit in Figure 1. %\ref{fig:data}.

%%%%%% Figure 2 %%%%%%%%%%%%%%%%%
\begin{figure*}[t!]
%\picplace{5cm}
\centerline{\psfig{figure=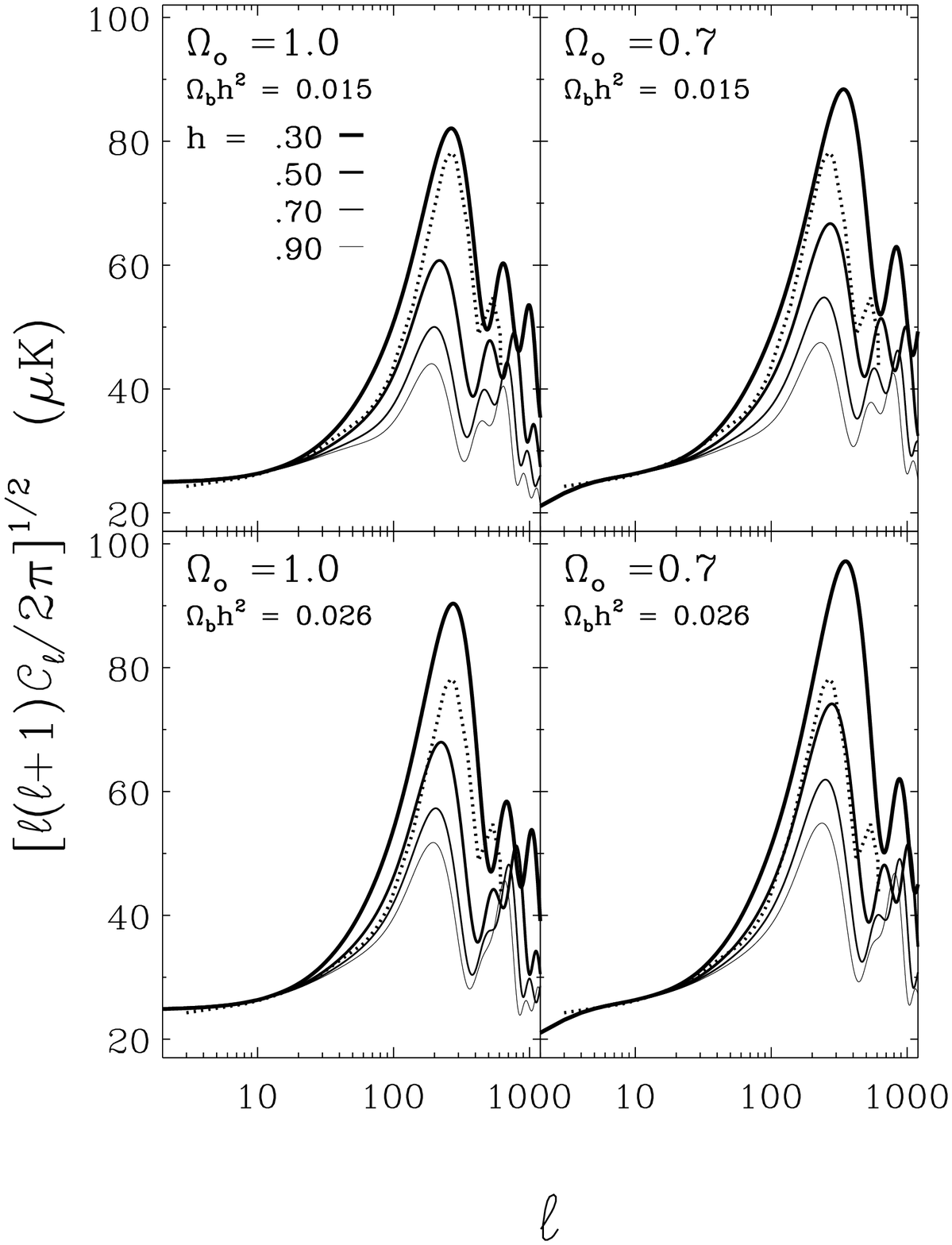,height=14cm,width=16.cm,bbllx=25pt,bblly=50pt,bburx=560pt,bbury=730pt}}
%\centerline{\psfig{figure=cl4ohbbn.ps,height=8cm,width=9.cm,bbllx=25pt,bblly=80pt,bburx=560pt,bbury=680pt}}
%\centerline{\psfig{figure=cl4ohn1.ps,height=10.0cm,width=9.6cm,bbllx=10pt,bblly=120pt,bburx=594pt,bbury=690pt}}
%\centerline{\psfig{figure=hob1.ps,height=9cm,width=\hsize,bbllx=10pt,bblly=120pt,bburx=594pt,bbury=690pt}}
%%%BoundingBox: 28 70 594 636
%
\caption{CMB power spectra showing the influence of $\oo$, $\obh$ and $h$.
These models are for $n=1$, $\Q=17\:\mu$K.
The dotted line represents the data. It is the same in all panels and is the same as in 
Figure 1.   %\protect\ref{fig:data}.
The peak amplitude $A_{peak}$ depends strongly on $h$ but also on $\obh$ and $\oo$.
The $\ell$ value of the peak, $\ell_{peak}$, is $\oo$ dependent but also
mildly $h$ dependent.
In the lower right panel $\oo=0.7$, $h=0.50$, $\obh = 0.026$ fits the data quite well
and is very close to the best-fit model for $n=1$ models (Figure 3).\label{cl4hobbn}}
\label{fig:cl4ohbbn}
\end{figure*}    %(cl4ohbbn.pro)
%%%%%%%%%%%%%%%%%%%%%%%%%%%%%%%%%%%%%%%%%%%%%%%%%%%%%%%%%%%%%%%%%%%%%%%%%%%%%%%%%%%%%%%%%%%%%%%%%%%%%%%

Essentially, we are trying to find the parameters of the model that looks most like 
the dotted line in Figure 1.   %\ref{fig:data}.
Figure 2     %\ref{fig:cl4ohbbn} 
is an example of some of the model power spectra tested.
For each point in the 4-D parameter space we obtain a 
value for $\chi^{2}(h, \oo, n, \Q)$.
The parameter values at the minimum value ($\chi^{2}_{min}$) are the
best-fit parameters. 
The error bars we quote for each parameter are from 
the maximum and minimum
parameter values within the 4-D surface which satisfies
$\chi^{2}(h,\Omega, n, Q_{10}) = \chi^{2}_{min} + 1$.
To display the result we project this surface onto
the two dimensions of our choice.
The $\chi^{2}$ calculation is described in more detail in papers 1 and 2.

%%%%%% Figure 3 %%%%%%%%%%%%%%%%%
%\figcaption[hon1qfree101526.ps]{
%\noindent Figure 3\\
%%%%%%%%%%%%%%%%%%%%%%%%%%% h-Omega   n=1, Q=free  convopen.pro  convijklo.pro  %%%%%%%%%%%%%%%%%%%%%%%%%%%
\begin{figure}[t!]
%\vskip-1.0
%\centerline{\epsfig{file=newfig3.ps,height=2.0cm,width=3.0cm,bbllx=25,bblly=110,bburx=540,bbury=610}}
\centerline{\psfig{figure=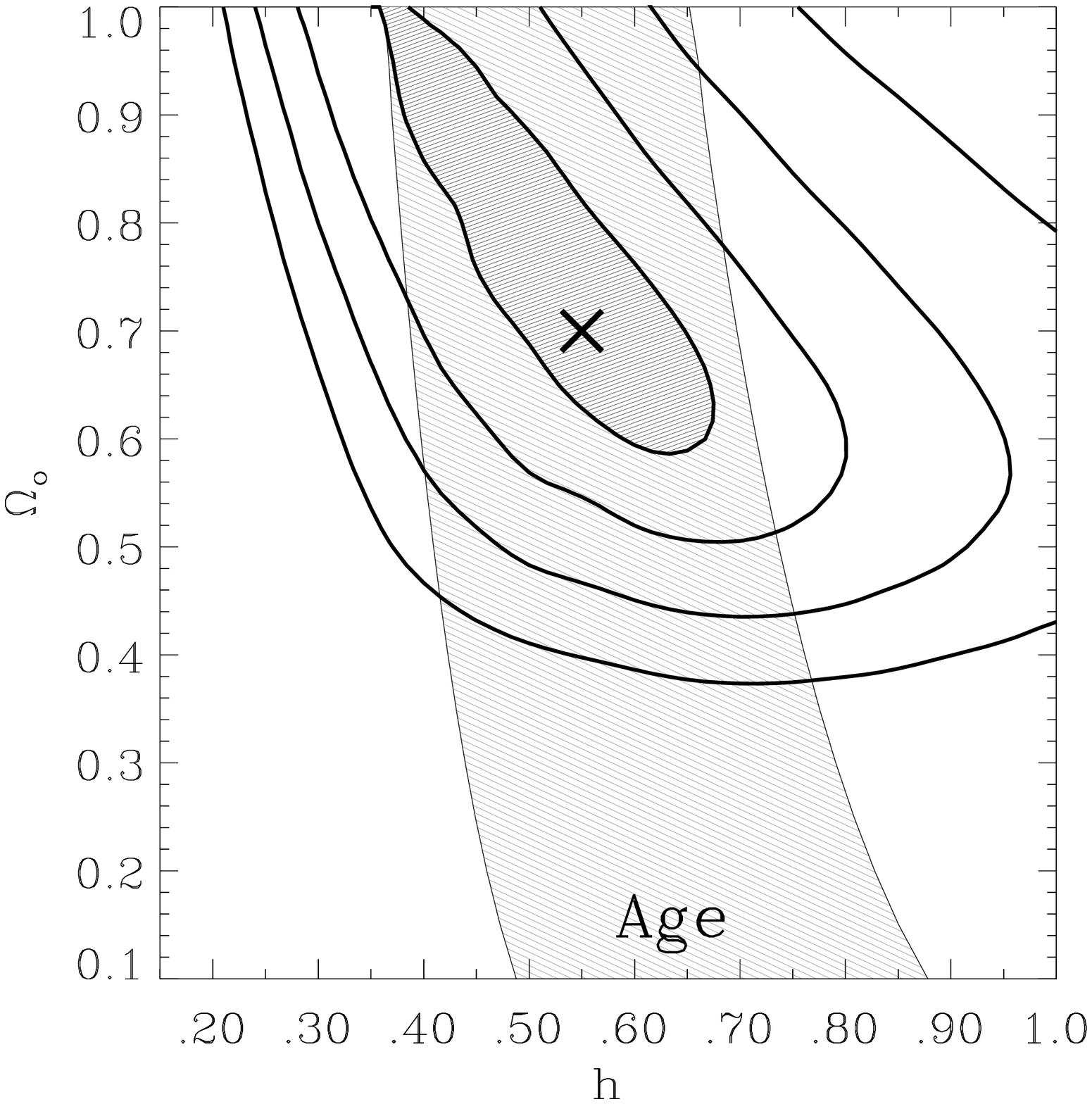,height=9.0cm,width=9.0cm,bbllx=25pt,bblly=110pt,bburx=540pt,bbury=610pt}}
%\centerline{\psfig{figure=hon1qfree101526.ps,height=8cm,width=9.cm,bbllx=25pt,bblly=110pt,bburx=540pt,bbury=610pt}}
%\centerline{\psfig{figure=hon1qfreebbn1526i.ps,height=8cm,width=9.6cm,bbllx=28pt,bblly=120pt,bburx=594pt,bbury=690pt}}
%\centerline{\psfig{figure=hon1q17L10.ps,height=10.0cm,width=9.cm,bbllx=28pt,bblly=120pt,bburx=594pt,bbury=690pt}}
%\centerline{\psfig{figure=hon1q17L10i.ps,height=8cm,width=9.6cm,bbllx=28pt,bblly=120pt,bburx=594pt,bbury=690pt}}
%\centerline{\psfig{figure=hn3.ps,height=9cm,width=\hsize,bbllx=10pt,bblly=120pt,bburx=594pt,bbury=690pt}}
%%%BoundingBox: 28 42 538 608
%\vskip-1cm
\caption{
Likelihood contours in the $h - \oo$ plane. 
We condition on $n=1$ while $\Q$ and $\obh$ are free to take on the value
that minimizes the $\chi^{2}$ value at that point.
The four contours correspond to $\chi^{2}_{min} + \Delta \chi^{2}$ where
$\Delta \chi^{2} = [ 1, 4, 9, 16]$ (See Section 2.1).   
The best-fit parameters are $h = 0.55^{+0.13}_{-0.19}$ and 
$\oo =0.70$ with $\oo > 0.58\; (\sim 1\sigma$).
The lightly shaded region represents the age constraint 
$10  < t_{o}  < 18$ Gyr.\label{hon1qfree}}
\end{figure}  %GET RID OF BAR ON TOP.(convopen.pro)

%%%%%% Figure 4 %%%%%%%%%%%%%%%%%
%\figcaption[honfreeqfree101526.ps]{
\begin{figure}[t!]
%\picplace{5cm}
%\centerline{\psfig{figure=newfig4.ps}}
\centerline{\psfig{figure=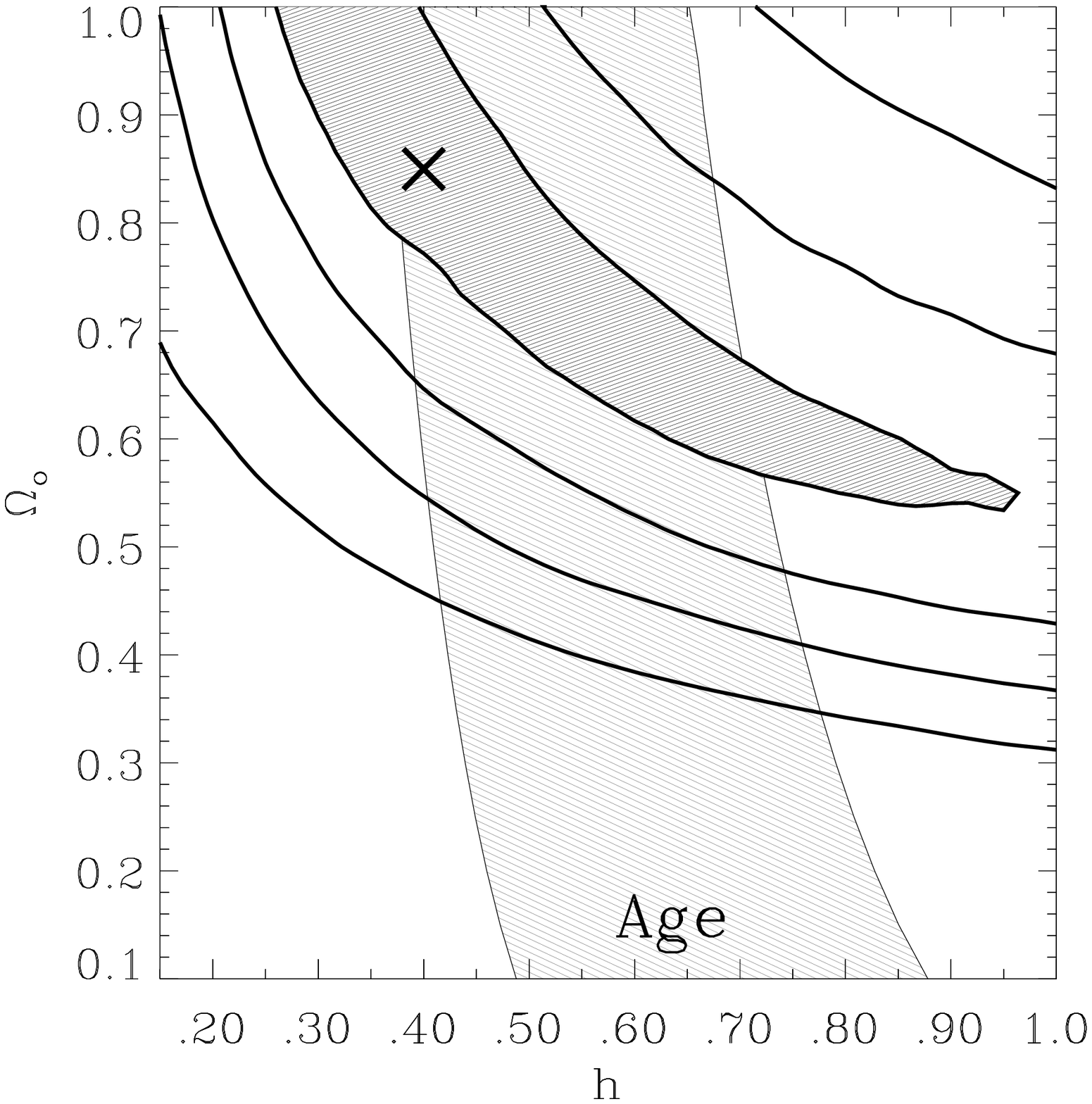,height=9.0cm,width=9.0cm,bbllx=25pt,bblly=110pt,bburx=540pt,bbury=610pt}}
%\centerline{\psfig{figure=honfreeqfree101526.ps,height=8.cm,width=9.cm,bbllx=25pt,bblly=110pt,bburx=540pt,bbury=610pt}}
%\centerline{\psfig{figure=honfreeqfreeL10.ps,height=18.0cm,width=18.cm,bbllx=10pt,bblly=120pt,bburx=594pt,bbury=690pt}}%\centerline{\psfig{figure=honfreeqfree1526i.ps,height=8.cm,width=9.6cm,bbllx=10pt,bblly=120pt,bburx=594pt,bbury=690pt}}
%\centerline{\psfig{figure=hn3.ps,height=9cm,width=\hsize,bbllx=10pt,bblly=120pt,bburx=594pt,bbury=690pt}}
%%%BoundingBox: 28 42 538 608
\caption{Same as previous figure except here we no longer condition on $n=1$.
Recall that our error bars are obtained from the projection of the
$\Delta \chi^{2} = 1$ contour onto an axis. 
Thus at $\sim 1\sigma$, $h$ is free to take on any value 
between $0.26$ and $0.97$.
At the minimum,
$h=0.40^{+0.57}_{-0.14}$ and $\oo = 0.85$ with $\oo > 0.53$.
The elongated $\Delta \chi^{2} = 1$ contour means that $h$ and $\oo$ 
are highly correlated. 
This correlation leads to a new constraint:
$0.45 < \oo h^{1/2} < 0.65$ which should be compared to the constraint on 
the same quantity from cluster baryonic fractions 
(see Figure 5     %\protect\ref{fig:nocmb101526} 
and Section \protect\ref{sec:clusters}).\label{honfreeqfree}}
\label{fig:honfreeqfree}
\end{figure}

Figure 3    %\ref{fig:hon1qfree} 
is the first in a series of contour plots 
which illustrates our results.
The four contours correspond to $\chi^{2}_{min} + \Delta \chi^{2}$  surfaces where
$\Delta \chi^{2} = [ 1, 4, 9, 16]$. 
The interpretation of these contours is not straightforward. 
The conditions under which these contours can be projected onto an axis 
yielding $1,\; 2,\; 3\;$ and $4 \sigma$ confidence intervals
is described in Press \etal (1992 p 690)
(see also Avni 1976).
These conditions are: i) the errors are normally distributed and 
ii) the model is linear in the parameters {\it or} 
that a linear approximation reasonably represents 
the models within the range of parameters of interest. 
deBernardis \etal (1997) find that the error bars are approximately normal.

Although the model power spectra are nonlinear in the parameters,
an approximation linear in the parameters may be able to represent the power 
spectrum near the $\chi^{2}$ minima.
For example, in Figure 2, several families of models
are plotted. The data have a maximum $\ell_{eff}$ value of $615$. The ability of the
$\chi^{2}$ to discriminate between models comes almost exclusively from the
data in the range $2 \lsim \ell_{eff} \lsim 430$.
The strongest nonlinearities in the models (typified by regions where 
the models do not
trace each other but overlap in complicated ways) are in the 
$\ell$ range to which the data is not sensitive. 
It could be argued that the relevant parts of the
models may be approximated by the first two terms
of a Taylor expansion of the power spectrum around the
best-fit parameter values. The accuracy of this linear approximation is a 
measure of the accuracy of the correspondence we would like to establish
between the $\chi^{2}_{min} + 1$ contour and the $1\sigma$ confidence interval.
The ellipticities of the contours in Figures 3, 4 and 7 are also a measure of 
the accuracy of the linear approximation; exactly linear models would give 
concentric, exact ellipses with identical orientations and position angles
and with semimajor axes in the ratios 1:2:3:4.
We conclude that the error bars that we derive from the 
$\chi^{2}_{min} + 1$ contours may be useful 
approximations to $1 \sigma$ confidence limits, but that the
$\chi^{2}_{min} + 4$, 9, and 16 contours are at best rough guides
to the 2, 3, and 4 $\sigma$ confidence intervals.
Work is in progress to quantify the accuracy of the linear approximation.
Bond, Jaffe \& Knox (1998) have done a preliminary
analysis comparing a $\chi^{2}$-minimization analysis of
flat-band estimates to a more complete pixel-based treatment. 
Their general conclusion is that our ``radical data compression method 
works...sort of'' since the minima found by the two techniques agree 
fairly well.

\subsection{Normalization and Definition of $\Q$}
\label{sec:q10}

We normalize in the middle of the COBE DMR data ($\ell=10$) 
rather than at the edge ($\ell = 2$) to reduce the  otherwise strong
correlation between the best-fit slope and normalization.
We parametrize the normalization at $\ell = 10$ using the symbol
$\Q$ {\it defined by}
\footnote{A CMB skymap can be written $\Delta T/T_{o}\;(\hat{n}) = \sum a_{\ell m} Y_{\ell m}(\hat{n})$ 
and its power spectrum is $C_{\ell} = \sum |a_{\ell m}|^{2}/(2\ell +1)$.
Note that here $C_{\ell}$ is not an ensemble average.}
\be
10(10+1)C_{10} = \frac{24\pi}{5}\frac{\Q^{2}}{T_{o}^{2}}.
\label{eq:c10}
\ee
Equation \ref{eq:c10} is simply a way to write $C_{10}$ with the added convenience
that for an $n=1$            %$n=1$ and $\oo = 1$, 
pure Sachs-Wolfe spectrum ($C_{\ell} \propto 1/(\ell(\ell + 1))$),
$\Q$ is equivalent to the power
spectrum normalizing quadrupole $Q_{rms-PS}$ (see Smoot \etal 1992).

\subsection{Saskatoon Calibration}
\label{sec:sk}
We have used the new calibration (Leitch 1998) for the Saskatoon results 
whereby the nominal Saskatoon calibration (Netterfield \etal 1995, Netterfield \etal 1997)
is increased by $5\%$ with a correlated calibration uncertainty around this new 
value of $7\%$.
We treat the calibration of the 5 Saskatoon points as a nuisance parameter ``$u_{sk}$''
coming from a Gaussian distribution with a dispersion of $7\%$ (rather than the $14\%$
used in paper 2).
In this sense our error bars include an estimate of the Saskatoon calibration uncertainty.

In general the minimum $\chi^{2}$ fits prefer
$u_{sk} \approx 0.86$.
This can be understood quite easily by examining Figure 1.    %\ref{fig:data}.
The little boxes above and below the Saskatoon points are $\pm 7\%$ of the central values.
$u_{sk} = 0.86$ corresponds to $-14\%$. Moving all 5 Saskatoon points down
by $\sim 14\%$  gives the best agreement with the dotted line, representing all the data.
Thus $u_{sk} \approx 0.86$ is the preferred value.

There are 32 data points and in the most general case where all 5
parameters ($h$, $\oo$, $n$, $\Q$, $u_{sk}$)
are free, there are 27 degrees of freedom  (= 32 - 5).   
When both $h$ and $\oo$ are low and thus nominally $\oo < \ob$; we set $\ob = \oo$, 
thus creating purely baryonic models in the lower left corners of Figures
3 and 4.     %\ref{fig:hon1qfree} and \ref{fig:honfreeqfree}.
Computer limits restrict the number of discrete $\obh$ values we can test.
We have performed all calculations for each of 3 values of $\obh$: 
$\obh \in \{0.010, 0.015, 0.026\}$. Thus we have explored three 4-D slices
of parameter space. For completeness we have also
minimized with respect to $\obh$ in the same way we have for the other parameters,
but the $\obh$ minimization is restricted to only three discrete values.
The results from this highly discretized BBN range minimization
are indicated by ``$\;*\;$'' in the $\obh$ column of Table 2.

%%%%%%%%%%%%%%%%%%%%%%%%%%%%%%%%%%%%%%%%%%%%%%%%%%%%
\subsection{Physical Effects in Open Models}

Acoustic oscillations of the baryon--photon fluid at 
recombination produce peaks in the CMB power spectrum
around degree angular scales.
It is convenient to discuss power spectra in terms of 
the amplitude and the position of the first such peak: $A_{peak}$ and $\ell_{peak}$.
For example, the amplitude and position of the polynomial fit to the data
(dotted line in Figure 1)   %\ref{fig:data}) 
are $\ap = 77\;\mu$K and $\lp=260$.
For the physics of the acoustic peaks, see the pioneering work by Hu (1995) and Hu \& Sugiyama (1995a,
1995b).

In Figure 2       %\ref{fig:cl4ohbbn} 
we plot CMB power spectra to display the influence of $\oo$ and $h$ and $\obh$.
The dotted line is the same in each panel, is the same as in Figure 1    %\ref{fig:data} 
and represents the data. 
These models are for $n=1$, $\Q=17\:\mu$K.
In Figure 2     %\ref{fig:cl4ohbbn} 
we can see that
the peak amplitudes $A_{peak}$ depend strongly on $h$ and $\obh$.
Higher values of $\obh$ lead to larger Doppler peaks due to 
the enhanced compression caused by a larger effective mass (more baryons per photon) of 
the oscillating fluid.
For a given $\obh$, a small $h$ means high $A_{peak}$.
$\Q$ variations would raise and lower the entire curve while variations in the slope
$n$ would raise and lower $A_{peak}$.

In Figure 2,    %\ref{fig:cl4ohbbn}, 
when $h \downarrow$, $\ell_{peak} \uparrow$.
Similarly, as $\oo \downarrow$, $\ell_{peak} \uparrow$.
The $\oo$ dependence of $\ell_{peak}$ is a purely geometric effect. The more open the universe, the smaller the
angle subtended by the same physical size.
The main point of Figure 2     %\ref{fig:cl4ohbbn} 
is that
lowering $h$ and lowering $\oo$ both have the same effect of raising
$\ell_{peak}$.
In paper 1 and 2 we maintained that it was predominantly the position of the peak that
favored low $h$ in $\oo=1$ models. Here we see that lowering $\oo$ can raise
$\ell_{peak}$ to fit the data, hence $h$ does not have to be as low.
The behaviour of $\ap$ and $\lp$ can be tabulated as,\\

\noindent
{\baselineskip=1cm
{\small
\begin{tabular}{|l|llll|}
\hline
$A_{peak} \uparrow    $&$ h \downarrow $&$ \oo \downarrow$ (for $\oo \gsim 0.6$) &$ \obh \uparrow$&$n \uparrow$\\
\hline     
$\ell_{peak} \uparrow $&$ h \downarrow $&$ \oo \downarrow                       $&$              $&\\
\hline
\end{tabular}
}}\\  %small

\noindent 
which is to be read ``$\ap$ goes up when $h$ or $\oo$ go down 
or when $\obh$ or $n$ go up''.
See Hu, Sugiyama \& Silk (1997) for more details.

%%%%%%%%%%%%%%%%%%%%%%%%%%%%%%%%%%%%%%%%%%%%%%%%%%%%%%%%%%%%%%%%%%%%%%%%%%%%%%%%%%%%%%%%%%%%%%%%%%%%%%%%%%%%
\section{RESULTS FROM THE $h - \oo$ DIAGRAM}

The $h - \oo$ diagram is a convenient framework in which to explore and present
a combination of cosmological parameters.
The regions preferred by the CMB are shown in 
Figures 3 and 4.       %\ref{fig:hon1qfree} and \ref{fig:honfreeqfree}. 
The results from these figures are given in the first two
sections of Table 2 which also contains the main results of this paper for $h$, $\oo$,
$n$ and $\Q$. 
For each result, the conditions under which it was obtained
are listed and these conditions are relaxed as we move from the top to the bottom.
Table 2 also lists $\chi^{2}$ values and the corresponding probabilities
$P(\chi^{2} <)$ of obtaining $\chi^{2}$ values smaller
than the values actually obtained, 
under the assumption that the errors on the data points are Gaussian.
$\chi^{2}$ values and probabilities are discussed in Section \ref{sec:chi}.

\subsection{$h$ Results}

For $\oo = 1$ we get $h=0.33 \pm 0.08$, which is the same low $h$ value 
we obtained in paper 2.
% (see the $1\sigma$ region's overlap with the BBN band 
%in Figure 3 of paper 2).
The new Saskatoon calibration and the new data used here do not change our
previous result.

In Figure 3   %\ref{fig:hon1qfree} 
we present the likelihood contours in the $h - \oo$ plane for $n=1$.
The minimum is at $h = 0.55^{+0.13}_{-0.19}$.
The minimum $\chi^{2}$ value and the probability of obtaining a smaller value for that model are
$\chi^{2}_{min}=21.4$ and $P(\chi^{2} <) = 23.4\%$ respectively. Thus the fit is ``good''.

Figure 4   %\ref{fig:honfreeqfree} 
is the same as Figure 3   %\ref{fig:hon1qfree}
except we no longer condition on $n=1$.
The best-fit $h$ value stays low but higher and lower $h$ values are now 
acceptable at $\sim 1\sigma$.
Thus, uncertainty in $n$ plays an important role in the inability of CMB data to determine $h$.
The banana-shaped $\Delta \chi^{2} = 1$ contour can be projected onto 
an axis to yield an approximation to a $1\sigma$ confidence interval 
around the best-fit value. Thus, $0.26 < h < 0.97$ or 
$h=0.40^{+0.57}_{-0.14}$.

A favored open model with $\oo = 0.3$ and
$h = 0.70$ is more than $\sim 4 \sigma$ from the CMB data
best-fit model and can be rejected based on goodness of fit at the
99\% confidence level. 
In contrast to our previous $\oo = 1$ results,
allowing $\oo < 1$ permits much larger $h$ values and there is no longer
a disagreement with more direct local measurements of $h$.
 
\subsection{$\oo$ Results}
Our $\oo$ results are also given in Table 2.
In Figure 3    %\ref{fig:hon1qfree} 
($n=1$) we obtain
$\oo =0.70$ and set a lower limit $\oo  > 0.58$.
We obtain no upper limit because we were unable to test $\oo > 1$ models.
In Figure 4   %\ref{fig:honfreeqfree} 
we obtain
$\oo =0.85$ with a lower limit of $\oo  > 0.53$ at $\sim 1\sigma$ 
and $\oo > 0.43$ at $\sim 2\sigma$.
Thus the CMB can place important constraints on these models.
 
If we assume that $h\approx 0.65 \pm 0.15$ (as indicated by local $h$ measurements) 
then the CMB data prefer a density in the range 
$0.54 \lsim \oo \lsim 0.84$ and a power spectral slope
$0.96 \lsim n \lsim 1.12$.

\subsection{A New Constraint on $\oo h^{1/2}$}

The variations in $n$ and $\obh$ permit the large
range of $h$ seen in Figure 4.   %    \ref{fig:honfreeqfree}.
 For example, for the highest values of $h$, 
$n \approx 1.18$ and $\obh = 0.026$ while for the lowest $h$ values
$n \approx 0.85$ and $\obh = 0.010$.
$n$ variations alone are not sufficient to permit very high $h$ values. For example
if we let $n$ be free but we condition on $\obh = 0.010$ then $h$ remains small: $h < 0.42$.
If we condition on $\obh = 0.015$ then $h < 0.51$.
$n$ and $\obh$ are keeping the peak amplitude fit correctly. 
High $h$ values suppress the peak height but this is compensated for by high $n$ and $\obh$.

The elongated banana-shaped $\Delta \chi^{2} = 1$ contour in Figure 4  %\ref{fig:honfreeqfree} 
means that $h$ and $\oo$ are anti-correlated.
In Figure 2     %\ref{fig:cl4ohbbn} 
we see that  $\ell_{peak} \uparrow$ when
$h \downarrow$ or $\oo \downarrow$.
Thus high $\oo$ go with low $h$ and low $\oo$ go with high $h$.
The $\Delta \chi^{2} = 1$ contour in Figure 4   %\ref{fig:honfreeqfree}
traces out this strong anti-correlation.
To get a constraint on two parameters simultaneously we need to look at the 
$\Delta \chi^{2} = 2.3$ contour. This can be described by $\oo h^{1/2} = 0.55 \pm 0.10$.
This should be compared to the constraint on 
the same quantity from cluster baryonic fractions (see Figure 5  %\ref{fig:nocmb101526}
and Section \ref{sec:clusters}).

%%%%%%%%%%%%%%%%%%%%%%%%%%%%%%%%%%%%%%%%%%%%%%%%%%%%%%%%
\section{NON-CMB CONSTRAINTS IN THE $h - \oo$ PLANE}
\label{sec:noncmbconstraints}

To view our results within a larger picture, we compare them
to other cosmological measurements and identify what the CMB constraints 
can add to this picture.
The independent non-CMB cosmological measurements are summarized below.
They are the same constraints used in paper 2 (with modifications described below)
and we now include local measurements of $h$.

Peacock \& Dodds (1994) made an empirical fit to the
matter power spectrum using a shape parameter $\Gamma$.
For $\oo \le 1$ models, $\Gamma$ can be written as (Sugiyama 1995)
\be
\Gamma = h\:\oo\: exp\left[-\ob\left(\frac{\oo+1}{\oo}\right)\right].
\ee
We adopt the $2\sigma$ limits of the empirical fit of 
Peacock \& Dodds (1994)(see also Liddle \etal 1996a) 
and include the $n$ dependence,
\be
\label{eq:gamlimits}
0.222 < \Gamma - 0.32\left(\frac{1}{n} - 1\right) < 0.293.
\ee
with the assumption that $0.8 \le n \le 1.2$.

\subsection{X-ray Cluster Baryonic Mass Fraction}
\label{sec:clusters}
Assuming that clusters are a fair sample of the Universe,
observations of the X-ray luminosity and the angular
size of galaxy clusters can be combined to constrain the quantity
$\frac{\ob }{\oo}h^{3/2}$.
We adopt the range $0.04 < \frac{\ob}{\oo} h^{3/2} < 0.10$
(White \etal 1993) with a central value of $0.06$ (Evrard 1997).
We include the uncertainty in the value of $\ob$ by
replacing $\ob$ with the BBN range $[ 0.010\:h^{-2}, 0.026\:h^{-2}]$
yielding the limits $0.10 < \oo\: h^{1/2} < 0.65$,
and the central value $\oo\: h^{1/2}= 0.25$.

%%%%%% Figure 5 %%%%%%%%%%%%%
%\figcaption[nocmbh65.ps]{
\begin{figure}[t!]
%\picplace{5cm}
\centerline{\psfig{figure=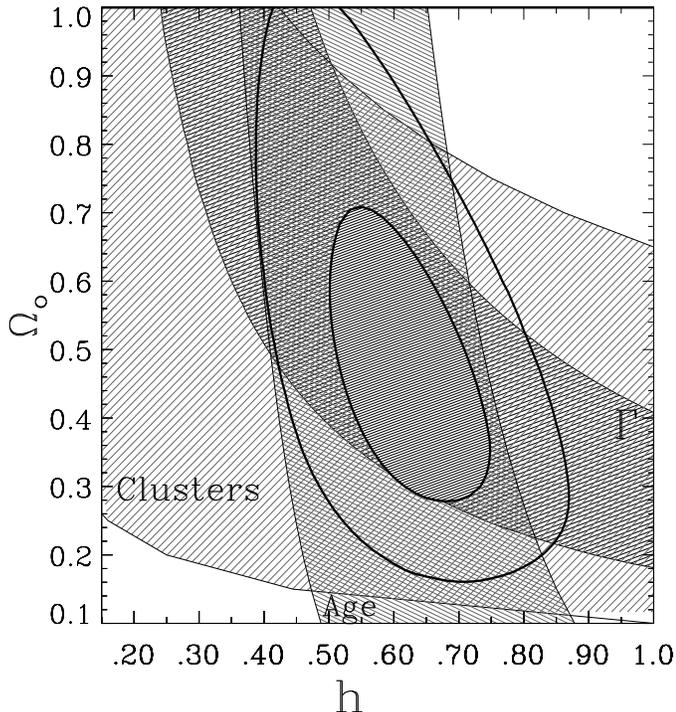,height=9.cm,width=9.cm,bbllx=25pt,bblly=110pt,bburx=540pt,bbury=610pt}}
%\centerline{\psfig{figure=nocmbh65.ps,height=8.cm,width=9.cm,bbllx=25pt,bblly=110pt,bburx=540pt,bbury=610pt}}
%\centerline{\psfig{figure=nocmb101526.ps,height=8.cm,width=9.cm,bbllx=25pt,bblly=100pt,bburx=540pt,bbury=610pt}}
%\centerline{\psfig{figure=nocmb15nogam.ps,height=8.cm,width=9.cm,bbllx=28pt,bblly=92pt,bburx=538pt,bbury=608pt}}
%%%BoundingBox: 28 42 538 608
\caption{This plot has no CMB information in it. The three bands are constraints from
three non-CMB cosmological measurements discussed in Section \protect\ref{sec:noncmbconstraints}.
The ``Age'' of the oldest stars in globular clusters: $10 < t_{o} < 18$ Gyr, 
the baryonic fraction in ``Clusters'' of galaxies:  $0.10 < \oo h^{1/2} < 0.65$ and
the matter power spectrum shape parameter ``$\Gamma$'': $0.169 < \Gamma < 0.373$.
The thick contours are approximate $1\sigma$ and $2\sigma$ regions from a
joint likelihood  of these three constraints with the added constraint from local 
measurements that $h=0.65 \pm 0.15$.
The results: $h=0.60^{+0.15}_{-0.10}$ and $\oo= 0.45^{+0.26}_{-0.17}$.
An uncertainty of the baryonic fraction $0.010 < \obh < 0.026$ has been included in the constraints.
In flat models the three constraints shown favor low values of $h \sim 0.40$ incompatible with local
measurements of $h$. 
In the open models considered here, this disagreement disappears.\label{nocmb101526}}
\end{figure}    %(nocmbo.pro)

%%%%%% Figure 6 %%%%%%%%%%%%%
%\figcaption[nocmball.ps]{
\begin{figure}[t!]
%\picplace{5cm}
\centerline{\psfig{figure=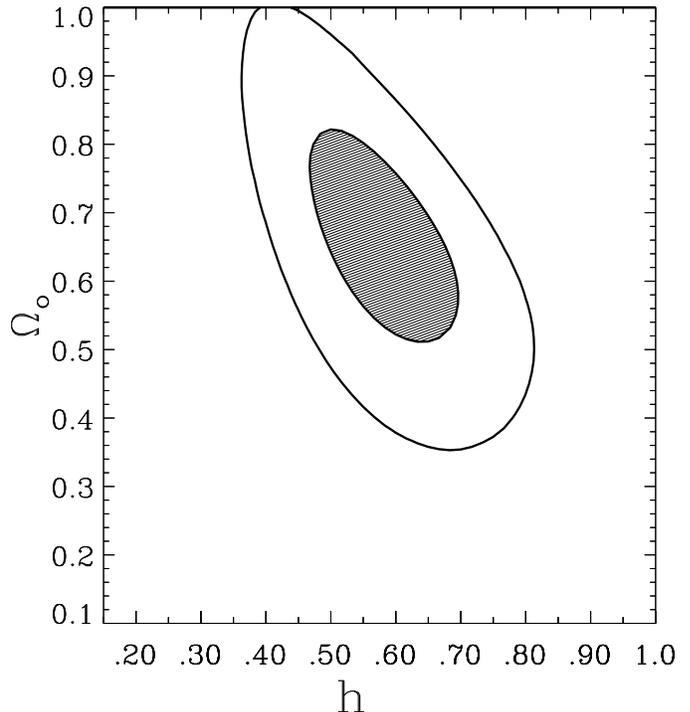,height=9.cm,width=9.cm,bbllx=25pt,bblly=110pt,bburx=540pt,bbury=610pt}}
%\centerline{\psfig{figure=nocmball.ps,height=8.cm,width=9.cm,bbllx=25pt,bblly=110pt,bburx=540pt,bbury=610pt}}
%%%BoundingBox: 28 42 538 608
\caption{Approximate $1\sigma$ and $2\sigma$ contours from a joint likelihood of the new CMB 
constraint $\oo h^{1/2} = 0.55 \pm 0.10$ with the four constraints shown in 
Figure 5.   %   \protect\ref{fig:nocmb101526}.
The results: $h=0.58 \pm 0.11$ and $\oo= 0.66^{+0.16}_{-0.15}$.
Popular $\oo = 1$ and $(\oo=0.3, h=0.70)$ models are 
$\approx 2\sigma$  from the best fit.\label{all}}
\end{figure}    %(nocmbo.pro)

%%%%%%%%%%%%%%%%%%%%%%%%%%%%
\subsection{Limits on the Age of the Universe from the Oldest Stars in
Globular Clusters}

Although the determinations of the age of the oldest stars in
globular clusters  are $h$- and $\oo$-independent, they
do depend on the distance assigned to the globular clusters of
our Galaxy. 
In paper 2 we used  $11 < t_{o} < 18$ Gyr with a central 
value of 14 Gyr. We now adopt $10 < t_{o} < 18$ Gyr with a central 
value of 13 Gyr because the recent Hipparcos 
recalibration of the local distance ladder
increases the distance to the globular clusters (Feast \& Catchpole 1997, 
Gratton \etal 1997, Reid 1997).
This lowers the inferred ages by about $\sim 5 - 10\%$ depending on what values were used 
in the calculation of the globular cluster distances.

Age determinations are $\oo$ and $\ol$ independent but converting them
to limits on Hubble's parameter depends on $\oo$ and on our $\ol=0$ assumption.
For a flat Universe with $\ol =0$ and $t_{o}$ expressed in Gyr,
$h = (6.52/t_{o})$. In an open universe 
\be
h = \left(\frac{9.78}{t_{o}}\right)\!
\left[\frac{1}{1-\oo}-\frac{\oo}{2(1-\oo)^{3/2}}
\cosh^{-1}\!\left(\frac{2-\oo}{\oo}\right)\right],
\label{eq:age}
\ee
where the term in square brackets is 2/3 for $\oo = 1$ and goes to 1 as  $\oo \rightarrow 0$.
The constraint $10  < t_{o} < 18$ Gyr,
inserted into Equation \ref{eq:age}, provides the ``Age'' constraint on $h$ used in 
Figures 3, 4 and 5.  %\ref{fig:hon1qfree}, \ref{fig:honfreeqfree} and  \ref{fig:nocmb101526}.

\subsection{Summary of Constraints Used}

The constraints we adopt from cluster baryonic fraction, 
the ages of the oldest stars in globular clusters,
the matter density power spectrum shape parameter $\Gamma$ and
local measurements of $h$ are,

{\baselineskip=.8cm
{\small
\begin{tabular}{llcccr}
Clusters  &$  0.10    $&$<$&$  \oo\;h^{1/2}  $&$<$&$ 0.65   $\\% &$ 0.25$\\
Age [Gyr] &$  10      $&$<$&$     t_{o}      $&$<$&$  18    $\\% &$ 13$ \\
$\Gamma  $&$  0.169   $&$<$&$   \Gamma       $&$<$&$ 0.373  $\\% &$ 0.25$\\ 
Hubble    &$  0.50    $&$<$&$       h        $&$<$&$ 0.80   $\\% &$ 0.65$\\ 
\end{tabular}
}}\\ %small 

\noindent where the respective central values adopted are
$\oo\;h^{1/2}=0.25$, $t_{o} = 13$ Gyr, $\Gamma = 0.25$ and $h=0.65$.
%******************************************************

The first three constraints are illustrated by the three bands in
Figure 5.   %    \ref{fig:nocmb101526}.
The $1\sigma$ and $2\sigma$ regions from an approximate joint likelihood
of all four constraints are also shown (see paper 2, Section 4.5 for details).
The $1\sigma$ region yields: $h=0.60^{+0.15}_{-0.10}$ and $\oo= 0.45^{+0.26}_{-0.17}$.
If we consider only the first three constraints the result is
$h=0.60^{+0.18}_{-0.21}$ and $\oo= 0.45^{+0.43}_{-0.17}$.
Thus, in $\oo \le 1$ models, there is good agreement between
the first three constraints and local measurements of $h$. This was not the case for
the $\oo = 1$ universes tested in paper 2 where 
the first three constraints favored lower values; $h \approx 0.40$
(notice in Figure 5     %\ref{fig:nocmb101526}
that for $\oo = 1$, $h \approx 0.40$ is preferred).
 
An uncertainty of the baryonic fraction of $0.010 < \obh < 0.026$ has been included in both the
cluster and $\Gamma$ constraints.
We have also made a figure analogous to Figure 5    %\ref{fig:nocmb101526} 
but with
a smaller BBN uncertainty around a higher value, $0.022 < \obh < 0.026$.
For this case, the lower limits of the ``Cluster'' and ``$\Gamma$'' bands are raised, thus narrowing
the $1 \sigma$ region.

%%%%%%%%%%%%%%%%%%%%%
\subsection{Comparison of CMB and Non-CMB Constraints in the $h-\oo$ Plane}

What does the CMB add to the larger picture provided by these 
non-CMB measurements?

$\bullet$ Overall consistency:
A superposition of Figures 4 and 5     %\ref{fig:honfreeqfree} and  \ref{fig:nocmb101526}
shows that the $\sim 1 \sigma$ regions of CMB and  non-CMB overlap for $0.52 \lsim h \lsim 0.67$
and $0.58 \lsim \oo \lsim 0.71$.

$\bullet$ More detailed consistency:
The region of overlap of the first three constraints and the CMB  is in agreement with 
local measurements of $h$.
This agreement between CMB, three independent cosmological measurements
and local $h$ measurements is non-trivial; in paper 2, although we had agreement
between the CMB and  three independent cosmological measurements, the agreement was at $h \sim 0.35$
and did not agree with local $h$ measurements.

$\bullet$ New constraint:
We find a tight new model-dependent constraint $\oo h^{1/2}= 0.55 \pm 0.10$ 
which favors the higher values of the cluster constraint on this same quantity.

$\bullet$ Preference for high $\obh$:
Inside the $\Delta \chi^{2} = 1$ contour (except for a small region to the left of the best fit)
of Figure 4,   %\ref{fig:honfreeqfree}, 
the $\chi^{2}_{min}$ is for $\obh = 0.026$.
The consistency between the non-CMB constraints and the CMB constraints
is stronger for higher values of $\obh$.
This improved consistency and slightly better fit indicates that high $\obh$ is
preferred, lending some support to Tytler \etal (1997) values.

$\bullet$ New argument for $n \sim 1$:
Figure 3,   %\ref{fig:hon1qfree}
where $n = 1$, has a minimum inside the joint likelihood $1\sigma$ contour of Figure 5.
In this sense it is more consistent with the combined constraints of Figure 5 than are
the results of Figure 4.      % %\ref{fig:nocmb101526} 
We can turn the argument around and say that the non-CMB constraints favor 
$n \sim 1$ based on this consistency.

$\bullet$ More precise combined constraint:
Combining the CMB constraint on $\oo h^{1/2}$ 
with the non-CMB contours in Figure 5   %\ref{fig:nocmb101526}
we obtain: $h=0.58 \pm 0.11$ and
$\oo = 0.66^{+0.16}_{-0.15}$ (see Figure 6).   %\ref{fig:all}).

$\bullet$ Rejection of a favored model:
The $\oo = 0.3$, $h = 0.70$ model is acceptable to the non-CMB measurements but is more 
than $\sim 4\sigma$ away from the best-fit CMB model in Figure 4   %\ref{fig:honfreeqfree}
and can be rejected based on goodness of fit at the $99\%$ CL.

Liddle \etal (1996a) have examined open models with $\ol = 0$.
They consider the shape parameter $\Gamma$, bulk flows,
the abundance of clusters and the abundance of Ly-$\alpha$ systems and the
age of the Universe. 
They find a good fit for $\oo > 0.35$ and an alarmingly low
$h$ value for $\oo \sim 1$. Assuming $h > 0.6$ (as indicated by many recent measurements)
they get $0.30 < \oo < 0.60$.
If we assume $h > 0.6$ we obtain $0.53 < \oo < 0.75$ from the CMB analysis,
and $0.3 < \oo < 0.70$ from the first three non-CMB constraints.
Thus we find consistent but slightly higher allowed intervals for $\oo$.

%%%%%%%%%%%%%%%%%%%%%%%%%%%%%%%%%%%%%%%%%%%%%%%%%%%%%%%%%%
%%%%%% Figure 7 %%%%%%%%%%%%%
%\clearpage
%\figcaption[nqhfreeofree101526.ps]{
\begin{figure}[t!]
%\picplace{5cm}
\centerline{\psfig{figure=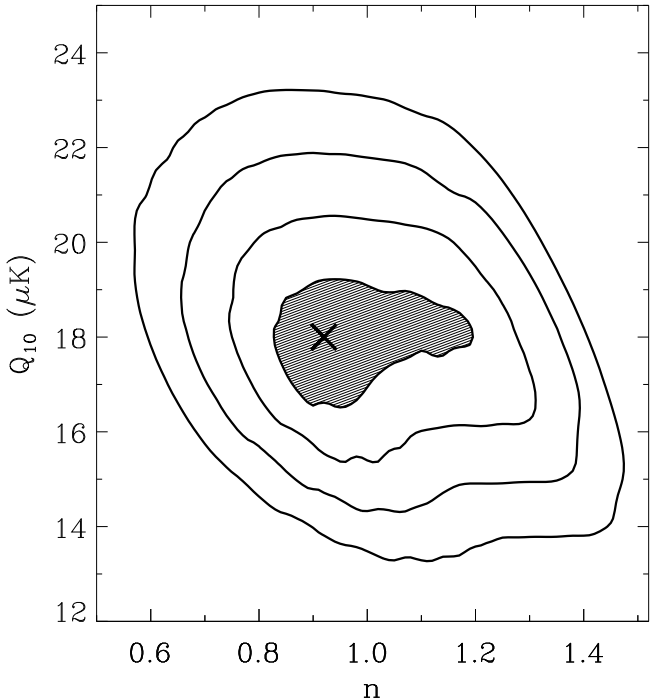,height=9.0cm,width=9.cm,bbllx=25pt,bblly=110pt,bburx=540pt,bbury=610pt}}
%\centerline{\psfig{figure=nqhfreeofree101526.ps,height=8.0cm,width=9.cm,bbllx=25pt,bblly=110pt,bburx=540pt,bbury=610pt}}
%\centerline{\psfig{figure=nqhfreeofree101526.ps,height=8.0cm,width=9.cm,bbllx=28pt,bblly=92pt,bburx=538pt,bbury=608pt}}
%\centerline{\psfig{figure=nqhfreeo1i.ps,height=8.0cm,width=9.cm,bbllx=28pt,bblly=92pt,bburx=538pt,bbury=608pt}}
%%%BoundingBox: 28 42 538 608
%
\caption{Likelihood contours in the $n - \Q$ plane with $h$ and $\oo$ as free parameters.
The minimum is $n=0.91^{+0.29}_{-0.09}$, $\Q=18.0^{+1.2}_{-1.5}\;\mu$K.
Notice that there is no strong correlation between $n$ and $\Q$ as there is
between $n$ and $Q_{rms-PS}$ (e.g. Lineweaver 1994).\label{nqhfreeofree}}
\end{figure}

\section{RESULTS FOR  $n$, $\Q$, $A_{peak}$ AND $\ell_{peak}$}

\subsection{Results for $n$ and $\Q$}

Our $n$ results are listed in Table 2.
Conditioning on $\oo = 1$ we get $n=0.91^{+0.10}_{-0.10}$. 
Figure 7    %\ref{fig:nqhfreeofree} 
displays our most general result in the $n -\Q$ plane
and yields $n=0.91^{+0.29}_{-0.09}$. 
Thus the minimum is unchanged and the error bars
increase slightly in this more general case. The best-fit value of $n$ is a robust result
in the sense that it does not change from the $\oo =1$ to the $\oo \le 1$ case.

For $\oo=1$ and $n=1$
we get  
$\Q = 17.0^{+1.4}_{-1.0} \;\mu$K.
For $\oo$ free and $n=1$ we obtain 
$\Q=18.0^{+1.4}_{-1.5} \;\mu$K.
Conditioning on $\oo = 1$ we obtain
$\Q=17.5^{+1.1}_{-1.1} \;\mu$K.
And finally with all other parameters free 
we get $\Q=18.0^{+1.2}_{-1.5} \;\mu$K (Figure 7).   %\ref{fig:nqhfreeofree}).
We can also express this normalization in terms of $C_{10}$,
\be
10^{11}C_{10} = 0.597 \left(\frac{\Q}{18.0\; \mu K}\right)^{2}.
\ee
This normalization at the best-fit values for $h$, $\oo$, $n$ and $\obh$
should be compared to the slightly higher, more general (but COBE DMR only)
Bunn \& White (1997) normalization
which is a function of the first and second derivatives
of the power spectrum at $\ell = 10$.

\subsection{Results for $A_{peak}$ and $\ell_{peak}$}

We can get a rough idea of the values of $\ap$ and $\lp$  preferred
by the data from the sixth order polynomial fit shown in Figure 1.   %\ref{fig:data}.
This yields $\ap \approx 77\;\mu$K and  $\lp \approx 260$.
We can also make a more careful model-dependent estimate of $\ap$ and $\lp$ by
looking at the power spectrum of the best-fit model in Figure 4   %\ref{fig:honfreeqfree}
and by examining the power spectra from models along the edge of 
the $\Delta \chi^{2} = 1$ contour.
The power spectrum of the best-fit model gives us central values for $\ap$ and $\lp$ while 
the power spectra of the models along the edge of the 
$\Delta \chi^{2} = 1$ contour yield error bars on these central values.
The result is $\ap = 76^{+3}_{-7} \;\mu$K and $\lp = 260^{+30}_{-20}$.
It should be remembered that these results depend on the correctness of the
family of models we are considering ($\oo \le 1$, $\ol = 0$).

\subsection{$\chi^{2}$ and Probabilities P$(\chi^{2} <)$}
\label{sec:chi}

The minimum $\chi^{2}$ values are given in Table 2 along with
the probability of obtaining smaller values under a Gaussian assumption for the
errors on the flat-band power estimates.
The range of minimum-$\chi^{2}$ values and their corresponding probabilities are
 $[ 20.9 < \chi^{2} < 24.6]$ and $[18.2 < P < 35.2]$.
These $\chi^{2}$ values are ``good'' and border on ``too good''.
The highest $\chi^{2}$ values and the highest probabilities are when we condition on 
$h=0.50$ with $\oo = 1$ giving substantially worse fits than $h$ free.
The lowest probabilities are when we condition on $\oo = 1$ and $\obh = 0.026$.

We have added the calibration uncertainty in quadrature to the statistical
error bars on the flat-band power estimates.
If we were more conservative we would add them  linearly.
In this case the $\chi^{2}$ values would be even lower and the fits even better, 
i.e., ``too good''.

Figure 2    %\ref{fig:cl4ohbbn}
shows how the CMB power spectra vary as a function of  
$\obh$. 
In Table 2 we have included results for each value of $\obh$ separately.
The $\obh=0.010$ minima have the highest $\chi^{2}$ values while
the $\obh=0.026$ minima have the lowest and are thus identical to 
the $\chi^{2}$ of the restricted minimization with respect to $\obh$ (see Section \ref{sec:sk}).

$\oo = 1$ models are a subset of the models examined here.
The differences between the $\oo=1$ results reported here and those
reported in paper 2 are small and can be understood by the three
differences in the analysis. In the present work 
i) we only look at three discrete values of of $\obh$
(in paper 2 we explored the $h - \ob$ plane),
ii) we include 5 more data points,
iii) we use a Saskatoon calibration $5\%$ higher with a smaller uncertainty.

%%%%%%%%%%%%%%%%%%%%%%%%%%%%%%%%%%%%%%%%%%%%%%%%%%%%%%
\section{DISCUSSION AND SUMMARY}

\subsection{Review of Results}

We use CMB flat-band power estimates to obtain constraints on $h$, $\oo$, $n$ and $\Q$ 
in the context of $\oo \le 1$ CDM models of the Universe.
Conditioning on $\oo = 1$ we obtain $h= 0.33\pm 0.08$.
Allowing $\oo$ to be a free parameter reduces the ability of the
CMB data to constrain $h$ and we obtain $0.26 < h < 0.97$ 
with the minimum at $h=0.40$. 
We obtain $\oo = 0.85$ and set a lower limit $\oo > 0.53$.
We find a strong correlation between acceptable
$h$ and $\oo$ values leading to the new CMB constraint
$\oo h^{1/2}= 0.55 \pm 0.10$.
We also obtain
$n=0.91^{+0.29}_{-0.09}$ and $\Q= 18.0^{+1.2}_{-1.5}\;\mu$K.
High baryonic models ($\obh \sim 0.026$) yield the best CMB $\chi^{2}$ fits
which also are more consistent with other cosmological constraints.
We find that a favored open model with $\oo = 0.3$ and
$h = 0.70$ is more than $\sim 4 \sigma$ from the CMB data
best-fit model and can be rejected at the 99\% CL based on goodness of fit.

\subsection{Consistency with Non-CMB Measurements}

In the flat CDM models of paper 2 we found that $h\sim 0.30$.
This value was consistent with four non-CMB constraints but
in disagreement with local measurements of $h$.
Considering $\oo \le 1$ models for $\obh$ fixed at $0.015$ we again 
find $h$ limited to values $\lsim 0.50$ and best-fit values of $\oo$ near 1.
It is not until we allow $\obh = 0.026$ that a much larger 
interval of $h$ is allowed at the $\sim 1\sigma$ level. For this most general case,
the results from the CMB, the same non-CMB constraints as used previously 
and local measurements of $h$ are all consistent with 
$h\approx 0.58$ and $\oo \approx 0.66$.

The $\sigma_{8}$ that corresponds to our best-fit model 
is $\sigma_{8} = 0.51^{+1.76}_{-0.20}$.
This is consistent with results from independent measurements which favor the
interval $[0.4, 0.9]$ (Viana \& Liddle 1996, Eke \etal 1996).
Our $\oo$ results are also broadly consistent with bulk flow measurements which yield
roughly $\oo > 0.4$ (Dekel 1997).

CMB constraints are independent of other cosmological measurements
and are thus particularly important. 
The fact that reasonable $\chi^2$ values are obtained means that
the current CMB data are consistent with inflation-based
$\oo \le 1$ CDM models for a broad range of $h$ values.
In the context of the models considered, the CMB results are consistent with
three other independent cosmological measurements and
are now also in agreement with local measurements of $h$.
This consistency was not present in $\oo = 1$ models.

\subsection{Review of Assumptions}
\label{sec:cond}

The results we have presented here are valid under the assumption
of inflation-based CDM models with Gaussian adiabatic initial conditions 
and with no cosmological constant. 
We have not considered early reionization scenarios or hot dark matter.
We have also not included any gravitational wave contributions which seem
to make the fits slightly worse without changing the
location of the best-fit parameters (Liddle \etal 1996b, Bond and Jaffe 1997).
With only scalar perturbations, deviations of the power spectrum from power-law behavior
is negligible in open models (Garcia-Bellido 1997). 
Supercurvature modes are not included in the power spectra models 
(Seljak \& Zaldarriaga 1996)
because they are in general important only for $\ell \lsim 5$ and thus difficult 
to measure because of cosmic variance.

It is possible that one or more of our basic assumptions is wrong,
or we could simply be looking at too restricted a region of parameter space. 
Topological defects may be the origin of structure. 
Using the same data and $\chi^{2}$-minimization analysis, we find
(Durrer \etal 1997) that several classes of
scalar-component-only  global topological defect models also produce acceptable fits to the data
although the goodness of fit of these models is not as good as the models we
consider here.
In other words, goodness-of-fit statistics from current CMB data 
have a slight preference for the inflation-based models we have considered
over the topological defect models we have considered.

\subsection{Future Improvements}

In addition to the  $h$, $\oo$, $n$, $\Q$ and $\obh$ considered here,
regions of a larger dimensional parameter space deserve further investigation
including $\ol$, $\oh$, $\ob$, early reionization parameters such as $z_{reion}$, 
tensor mode parameters $n_{T}$ and $T$, the inflaton potential $r$,
iso-curvature or adiabatic initial conditions and topological defect models with 
their additional parameters.

The fact that we obtain acceptable $\chi^{2}$ values in our 4-D parameter
space lends some support to the idea that we may be close to the right model.
If the Universe is not well described by these models then as the data improve,
work like this will show poor $\chi^2$ fits and other regions of parameter space will
be preferred.

To increase the parameter-constraining power of the measurements, 
observations need to be made in regions of $\ell$-space that have no or 
few measurements. 
In Figure 1    %\ref{fig:data} 
we can identify these regions:
$600 < \ell < 1200$, 
$20 < \ell < 50$ and  $180 < \ell < 400$.
More than a dozen on-going small-angular-scale experiments continue to fill in these 
gaps (Page 1997) as we 
await the more definitive MAP and Planck satellite results.

The improvement of non-CMB measurements will reduce the size of parameter space we need
to look at making the model power spectra computations more tractable. For example
if $\obh$ can be determined to be $\obh = 0.024 \pm 0.002$ as claimed by Tytler \etal (1997)
(or some other equally well-constrained value)
then a much smaller range of the $h-\ob$ plane needs to be examined and 
the range of $h$ allowed by the CMB analysis will be much narrower.
The indeterminacy of $n$, which seems to be measurable solely by the CMB
and whose error bar has a relatively large contribution from 
irreducible cosmic variance, will remain a dominant factor in the uncertainty of 
CMB parameter estimation.

We gratefully acknowledge the use of the Boltzmann code kindly
provided by Uros Seljak and Matias Zaldarriaga.
We benefited from useful discussions with Max Tegmark and Luis Tenorio. 
C.H.L. acknowledges
support from the North Atlantic Treaty Organization under a grant awarded in 1996.
D.B. is supported by the Praxis XXI CIENCIA-BD/2790/93 grant attributed by 
JNICT, Portugal.

%%%%%%%%%%%%%%%%%%%%%%%%%%%%%%%%%%%%%%%%%%%%%%%%%%%%%%%%%%%%%%%%%%%

%%%%%%%%%%%%%%%%%%%%%%%%%%%%%%%%%%%%%%%%%%%%%%%%%%%%%%%%%%%%%
\clearpage
%table is output by cl1.pro
{\scriptsize
\begin{table}[t]
\begin{center}
\caption{Data$^{a}$ Used in the $\chi^{2}$ Fits and Plotted in Figure 1}% and the \chisq Fits of Two Models$^{a}$}
\vspace{2pt}
%\begin{tabular}{|l|l|r|c|c|r|} \hline         % c:center,l:left,r:right
\begin{tabular}{|l|l|r|c|} \hline         % c:center,l:left,r:right
Experiment & reference  &$\ell_{eff}$& $\delta T_{\ell_{eff}}^{data} \pm \sigma ^{data}(\mu$K)\\%& $\chi^{2}(h=0.30)$ & $\chi^{2}(h=0.75)$\\
\hline
DMR1    &\protect\cite{hin96}&  3&$ 27.9_{-  4.0}^{+  5.6}$\\%&     0.042&     0.047\\
DMR2    &\protect\cite{hin96}&  7&$ 24.6_{-  2.8}^{+  3.6}$\\%&     0.803&     0.808\\
DMR3    &\protect\cite{hin96}& 14&$ 30.8_{-  3.1}^{+  3.4}$\\%&     0.281&     0.272\\
DMR4    &\protect\cite{hin96}& 25&$  1.2_{-  1.2}^{+ 14.4}$\\%&     4.511&     4.436\\
FIRS    &\protect\cite{gan94}& 10&$ 29.4_{-  7.7}^{+  7.8}$\\%&     0.008&     0.011\\
Tenerife&\protect\cite{gut97}& 20&$ 32.5_{- 8.5}^{+ 10.1}$\\%&     0.078&     0.086\\
SP91    &\protect\cite{gun95}& 60&$ 30.2_{-  5.5}^{+  8.9}$\\%&     1.564&     0.775\\
SP94    &\protect\cite{gun95}& 60&$ 36.3_{-  6.1}^{+ 13.6}$\\%&     0.137&     0.016\\
BAM     &\protect\cite{tuc97}& 74&$ 55.6_{- 15.2}^{+ 29.6}$\\%&     0.191&     0.877\\
Pyth1   &\protect\cite{pla97}& 87&$ 60_{- 13}^{+ 15}$\\%&     0.191&     0.877\\
Pyth2   &\protect\cite{pla97}&170&$ 66_{- 16}^{+ 17}$\\%&     0.369&     0.081\\
ARGO1   &\protect\cite{deb94}& 95&$ 39.1_{-  8.7}^{+  8.7}$\\%&     1.585&     0.343\\
ARGO2   &\protect\cite{mas96}& 95&$ 46.8_{- 12.1}^{+  9.5}$\\%&     0.117&     0.046\\
MAX GUM &\protect\cite{tan96}&138&$ 54.5_{- 10.9}^{+ 16.4}$\\%&     0.059&     0.317\\
MAX ID  &\protect\cite{tan96}&138&$ 46.3_{- 13.6}^{+ 21.8}$\\%&     0.313&     0.009\\
MAX SH  &\protect\cite{tan96}&138&$ 49.1_{- 16.4}^{+ 21.8}$\\%&     0.186&     0.002\\
MAX HR  &\protect\cite{tan96}&138&$ 32.7_{-  8.2}^{+ 10.9}$\\%&     5.602&     2.065\\
MAX PH  &\protect\cite{tan96}&138&$ 51.8_{- 10.9}^{+ 19.1}$\\%&     0.123&     0.099\\
Sk1     &\protect\cite{net97}& 86&$ 49.0_{-  5.0}^{+  8.0}$\\%&     0.104&     1.760\\
Sk2     &\protect\cite{net97}&166&$ 69.0_{-  6.0}^{+  7.0}$\\%&     0.513&     6.664\\
Sk3     &\protect\cite{net97}&236&$ 85.0_{-  8.0}^{+ 10.0}$\\%&     1.589&    14.668\\
Sk4     &\protect\cite{net97}&285&$ 86.0_{- 10.0}^{+ 12.0}$\\%&     0.946&    15.046\\
Sk5     &\protect\cite{net97}&348&$ 69.0_{- 28.0}^{+ 19.0}$\\%&     0.001&     1.340\\
MSAM    &\protect\cite{che96}&159&$ 40.7_{- 17.0}^{+ 30.5}$\\%&     0.001&     1.340\\
MSAM    &\protect\cite{che97}&159&$ 50_{- 9}^{+ 13}$\\%&     0.001&     1.340\\
MSAM    &\protect\cite{che96}&263&$ 44.4_{- 14.5}^{+ 23.9}$\\%&     0.001&     1.340\\
MSAM    &\protect\cite{che97}&263&$ 65_{- 10}^{+ 14}$\\%&     0.001&     1.340\\
CAT     &\protect\cite{sco96}&396&$ 51.8_{- 13.6}^{+ 13.6}$\\%&     0.823&     1.253\\
CAT     &\protect\cite{bak97}&422&$ 47.3_{- 6.3}^{+ 9.3}$\\%&     0.823&     1.253\\
CAT     &\protect\cite{sco96}&607&$ 49.1_{- 13.7}^{+ 19.1}$\\%&     0.207&     0.098\\
CAT     &\protect\cite{bak97}&615&$ 43.2_{- 10.1}^{+ 13.5}$\\%&     0.207&     0.098\\
OVRO    &\protect\cite{lei97b}&537&$ 56_{- 11}^{+ 14}$\\%&     0.823&     1.253\\
\hline
\end{tabular}
\end{center}
$^{a}$ 
CMB anisotropy detections reported in publications since 1994. The Netterfield
\etal (1997) points in Figure 1 are $5\%$ higher than these numbers due to the
Leitch (1998) recalibration.
See Lineweaver \etal (1997) and Lineweaver \& Barbosa (1998) for further details.\\ 

\end{table}
}   %end scriptsize
%%%%%%%%%%%%%%%%%%%%%%%%%%%%%%%%%%%%%%%%%%%%%%%%%%%%%%%%%%%%%%%%%%%%%%%%%%%%%%%%%%%%%%
\clearpage
%%%%%%%%%%%%%%%%%%%%%%%%%%%%%%%%%%%%%%%%%%%%%%%
%\footnotesize
%\tiny
{\scriptsize
\begin{table}
%\renewcommand{\arraystretch}{0.93}      %controls what [-4.0mm] does
%\footnotesize
%\footnotesep=0.8
\begin{center}
\caption{ Parameter Results}%% \\ [+2.0mm]
\begin{tabular}{|l l| c c c c c|r|}  \hline
\multicolumn{2}{|c|}{Result$^{a}$} &
\multicolumn{5}{c|}{ Conditions$^{b}$} &
\multicolumn{1}{|c|}{ $\chi^{2}\:(P(\chi^{2} <))^{c}$}  \\%% [-3.0mm]
        &                           &    h     & $\oo$ &  $n$  & $\Q(\mu$K)$^{d}$  & $\obh$& $(\;\%\:)$   \\%%[-1.0mm]
\tableline 
%%%%%\footnotesize
$H_{o}= $& $35^{+11}_{-4}$          &   --     &  1    &   1   &  free & \free  &$ 22.5(24.0)$\\%[-1.5mm]
         & $\mathbf{33^{+8}_{-8}}$  &   --     &  1    & free  &  free & \free  &$ 21.2(22.2)$\\%[-1.5mm]
         & $45^{+7}_{-13}$          &   --     & free  &   1   &  17   & \free  &$ 22.0(21.9)$\\%[-1.5mm]
         & $37^{+12}_{-6}$          &   --     & free  &   1   &  free & 0.010 &$ 22.6(24.7)$ \\%[-1.5mm]
         & $45^{+11}_{-13}$         &   --     & free  &   1   &  free & 0.015 &$ 22.2(22.8)$ \\%[-1.5mm]
         & $55^{+13}_{-14}$         &   --     & free  &   1   &  free & 0.026 &$ 21.4(19.5)$ \\%[-1.5mm]
         & $\mathbf{55^{+13}_{-19}}$&   --     & free  &   1   &  free & \free  &$ 21.4(23.4)$\\%[-1.5mm]
         & $30^{+12}_{-6}$          &   --     & free  & free  &  free & 0.010 &$ 21.6(24.3)$ \\%[-1.5mm]
         & $35^{+16}_{-5}$          &   --     & free  & free  &  free & 0.015 &$ 21.3(22.8)$ \\%[-1.5mm]
         & $40^{+57}_{-12}$          &   --    & free  & free  &  free & 0.026 &$ 20.9(21.1)$ \\%[-1.5mm]
         & $\mathbf{40^{+57}_{-14}}$ &   --    & free  & free  &  free & \free  &$ 20.9(25.4)$\\%%[-1.3mm]

\tableline
$\oo=$  & $0.85^{f}_{-0.16}$         &  free    &  --   &  1    &   17  & \free  &$ 22.0(21.9)$\\%[-1.5mm]
        & $1.00^{f}_{-0.28}$         &  free    &  --   &  1    &  free & 0.010 &$ 22.6(24.7)$ \\%[-1.5mm]
        & $0.85^{f}_{-0.21}$         &  free    &  --   &  1    &  free & 0.015 &$ 22.2(22.8)$ \\%[-1.5mm]
        & $0.70^{+0.28}_{-0.12}$     &  free    &  --   &  1    &  free & 0.026 &$ 21.5(19.6)$ \\%[-1.5mm]
        & $\mathbf{0.70^{f}_{-0.12}}$&  free    &  --   &  1    &  free & \free  &$ 21.5(23.8)$\\%[-1.5mm]
        & $1.00^{f}_{-0.21}$         &  free    &  --   & free  &  free & 0.010 &$ 21.6(24.3)$ \\%[-1.5mm]
        & $0.90^{f}_{-0.19}$         &  free    &  --   & free  &  free & 0.015 &$ 21.3(22.8)$ \\%[-1.5mm]
        & $0.85^{f}_{-0.32}$         &  free    &  --   & free  &  free & 0.026 &$ 20.9(21.1)$ \\%[-1.5mm]
        & $\mathbf{0.85^{f}_{-0.32}}$&  free    &  --   & free  &  free & \free  &$ 20.9(25.4)$ \\%%[-1.3mm]

\tableline
$n=$    & $1.03^{+0.08}_{-0.04}$     &   50     &  1    &  --   &  free & \free  &$ 24.6(35.2)$\\%[-1.5mm]
        & $0.91^{+0.11}_{-0.08}$     &  free    &  1    &  --   &  free & 0.010 &$ 21.6(20.1)$ \\%[-1.5mm]
        & $0.91^{+0.10}_{-0.06}$     &  free    &  1    &  --   &  free & 0.015 &$ 21.3(18.7)$ \\%[-1.5mm]
        & $0.91^{+0.07}_{-0.09}$     &  free    &  1    &  --   &  free & 0.026 &$ 21.2(18.2)$ \\%[-1.5mm]
        & $0.91^{+0.10}_{-0.10}$     &  free    &  1    &  --   &  free & \free  &$ 21.2(22.2)$\\%[-1.5mm]
        & $0.91^{+0.11}_{-0.08}$     &  free    & free  &  --   &  free & 0.010 &$ 21.6(24.3)$ \\%[-1.5mm]
        & $0.88^{+0.14}_{-0.05}$     &  free    & free  &  --   &  free & 0.015 &$ 21.3(22.8)$ \\%[-1.5mm]
        & $0.91^{+0.29}_{-0.09}$     &  free    & free  &  --   &  free & 0.026 &$ 20.9(21.1)$ \\%[-1.5mm]
        & $\mathbf{0.91^{+0.29}_{-0.09}}$& free & free  &  --   &  free & \free  &$ 20.9(25.4)$\\%%[-1.3mm]

\tableline
$\Q^{d}=$& $17.0^{+1.4}_{-1.0}$      &  free    &  1    &   1   &  --   & \free  &$ 22.5(24.0)$\\%[-1.5mm]
        & $17.5^{+1.2}_{-1.2}$       &   50     &  1    & free  &  --   & \free  &$ 24.6(35.2)$\\%[-1.5mm]
        & $18.0^{+0.9}_{-1.2}$       &  free    &  1    & free  &  --   & 0.010 &$ 21.6(20.1)$ \\%[-1.5mm]
        & $17.5^{+1.3}_{-1.1}$       &  free    &  1    & free  &  --   & 0.015 &$ 21.3(18.7)$ \\%[-1.5mm]
        & $17.5^{+1.2}_{-1.1}$       &  free    &  1    & free  &  --   & 0.026 &$ 21.2(18.2)$ \\%[-1.5mm]
        & $17.5^{+1.1}_{-1.1}$       &  free    &  1    & free  &  --   & \free  &$ 21.2(22.2)$\\%[-1.5mm]
        & $\mathbf{18.0^{+1.4}_{-1.5}}$&  free  & free  &   1   &  --   & \free  &$ 21.5(23.8)$\\%[-1.5mm]
        & $18.0^{+1.0}_{-1.3}$       &  free    & free  & free  &  --   & 0.010 &$ 21.6(24.3)$ \\%[-1.5mm]
        & $18.0^{+1.2}_{-0.5}$       &  free    & free  & free  &  --   & 0.015 &$ 21.3(22.8)$ \\%[-1.5mm]
        & $18.0^{+1.2}_{-1.5}$       &  free    & free  & free  &  --   & 0.026 &$ 20.9(21.1)$ \\%[-1.5mm]
        & $\mathbf{18.0^{+1.2}_{-1.5}}$&  free  & free  & free  &  --   & \free  &$ 20.9(25.4)$\\%%[-1.3mm]

\tableline
\end{tabular}\\
\end{center}
\tiny
${}^{a}$ parameter values at the the minimum $\chi^{2}$ values. 
The results cited in the abstract are in bold. The units of $H_{o}$ are $km\;s^{-1}\;Mpc^{-1}$\\
${}^{b}$ ``free'' means that the parameters were free to take on any values
within the discretely sampled ranges:
$ 0.15 \le h \le 1.00$, step size: 0.05, number of steps=18,
$ 0.1 \le \oo \le 1.0$, step size: 0.05, number of steps=19,
$ 0.49 \le n \le 1.51$, step size: 0.03, number of steps=35,
$ 12.0 \le \Q \le 25.0\:\mu$K, step size: 0.5 $\mu$K, number of steps: 26,
$ 0.010 \le \obh \le 0.026$, only three values: 0.010, 0.015 and 0.026.
Thus we have examined more than 900,000 models. See Section
\protect\ref{sec:cond} for more details about conditions.\\
${}^{c}$ Probability of obtaining a smaller $\chi^{2}$. There are 32 data points and the number of
degrees of freedom varies between 26 and 28.\\
${}^{d}$ $\Q=Q_{rms-PS}=Q_{flat}$ for pure Sachs-Wolfe, $n=1$ power spectra 
(see Section \protect\ref{sec:q10}). The units of $\Q$ are $\mu$K\\
${}^{e}$ Highly discretized minimization within the BBN range:  $\obh \in \{0.010, 0.015, 0.026\}$. See Section 2.3 for details.\\
${}^{f}$ The $\Delta \chi^{2}=1$ contour extends to values $\oo > 1$.\\

\end{table}
}   %end scriptsize
\normalsize
%%%%%%%%%%%%%%%%%%%%%%%%%%%%%%%%%%%%%%%%%%%%%%%%%%

%%%%%%%%%%%%%%%%%%%%%%%%%%%%%%%%%%%%%%%%%%%%%%%%%%%%%%%%%%%%%%%%%%%5

\end{document}